\documentclass[1p]{elsarticle}
\usepackage{bm}
\usepackage{amsmath,amssymb}
\usepackage{graphicx}

\newcommand{\BM}{\begin{pmatrix}}
\newcommand{\EM}{\end{pmatrix}}

\newcommand{\Imag}{\mathrm{Im}}

\newcommand{\bdgL}{\mathcal{L}}
\newcommand{\bdgM}{\mathcal{M}}
\newcommand{\tpartial}{\frac{\partial}{\partial t}}
\newcommand{\ve}{\varepsilon}
\newcommand{\bx}{x}
\newcommand{\by}{y}

\newcommand{\bs}{s}
\newcommand{\bt}{t}

\usepackage{color}

\begin{document}
\begin{frontmatter}

\title{Dynamical instability induced by zero mode \\ under symmetry breaking external perturbation}
%
\author[densi]{J.~Takahashi}
\ead{phyco-sevenface@asagi.waseda.jp}
\author[densi]{Y.~Nakamura}
\ead{nakamura@aoni.waseda.jp}
\author[densi]{Y.~Yamanaka}
\ead{yamanaka@waseda.jp}

\address[densi]{Department of Electronic and Photonic Systems, Waseda
University, Tokyo 169-8555, Japan}
\begin{abstract}
A complex eigenvalue in the Bogoliubov-de Gennes equations for a stationary Bose-Einstein condensate in ultracold atomic system indicates the dynamical instability of the system.
We also have the modes with zero eigenvalues for the condensate, called the zero modes, which originate from the spontaneous breakdown of symmetries.
Although the zero modes are suppressed in many theoretical analyses, we take account of them in this paper
and argue that a zero mode can change into one with a pure imaginary eigenvalue
by applying a symmetry breaking external perturbation potential. This emergence of a pure imaginary
mode adds a new type of scenario of dynamical instability to that characterized by
 complex eigenvalue of the usual excitation modes.
For illustration, we deal with two one-dimensional homogeneous Bose-Einstein condensate systems with a single dark soliton under a respective perturbation potential, breaking the invariance under translation, to derive pure imaginary modes.
\end{abstract}


\end{frontmatter}

\section{Introduction}
The Bose-Einstein condensation in ultracold neutral atomic 
systems~\cite{Cornell, Ketterle, Bradley}
is a macroscopic quantum phenomenon of indistinguishable bosonic particles.
The description of the Bose-Einstein condensate (BEC) based on the concept of a spontaneous breakdown of
symmetry is a success of quantum field theory.
Studies of the properties of BEC by means of quantum field theory will bring us a deeper understanding
of quantum physics. 

The instability of BEC is an interesting and challenging subject of the quantum
field theoretical study. Experimentally
the unstable phenomena belonging to the dynamical instability have been observed,
 such as BECs with quantized vortices~\cite{Shin}, in optical lattice~\cite{Fallani} and with solitons~\cite{Khaykovich}.
The dynamical instability originates from quantum fluctuations and should be distinguished from the Landau one which is due to energy dissipation.
Theoretically the dynamical instability is characterized by the existence of complex eigenvalues in the Bogoliubov-de Gennes (BdG) equations~\cite{Bogoliubov, deGennes, Fetter} which are
 obtained from linearizing the time-dependent Gross-Pitaevskii (TDGP) equation~\cite{Dalfovo}.
The field operators describing excitation modes are expanded in the complete set of the BdG 
eigenfunctions.
Our previous  work~\cite{Nakamura} has revealed that the emergence of complex 
eigenvalue for the excitation mode with non-zero real part of eigenvalue is attributed to 
the degeneracy between a positive-norm eigenfunction and a negative-norm one.

As a consequence of spontaneous breakdown of symmetry
and according to the Nambu-Goldstone theorem~\cite{NGtheorem},
there must be eigenfunctions of the BdG equations belonging to zero eigenvalues, called the zero modes.
These zero modes play such a crucial role that they create and retain the ordered states 
corresponding to the broken symmetries.
The simple BEC is interpreted as a spontaneous breakdown of the U(1) global gauge symmetry, 
and there appears the zero mode, called the $\theta$-zero mode.
Furthermore an additional symmetry may be broken spontaneously, and then one has another zero mode.
For example, the existence of a dark soliton in BEC implies the spontaneous breakdown of the translational symmetry for an originally homogeneous system, and there appears the zero mode 
related to the translational symmetry, called  the $x$-zero mode.
When the BdG equations are under consideration, it is necessary to introduce an additional adjoint mode 
corresponding to each zero one to make the complete set of the BdG 
eigenfunctions~\cite{Lewenstein, Matsumoto2, Dziarmaga}.
Although pairs of zero and adjoint modes  are ignored in many theoretical studies, mainly because of their infrared singular properties, we include them in our formulation of this paper.
Actually the pair of the $\theta$-zero mode and its adjoint one plays an essential role in the splitting process of a highly quantized vortex~\cite{Kobayashi}.

The purpose of this paper is to explore possible complex modes originating from zero modes.
First, as a general formulation, we consider a BEC system with an arbitrary number of zero modes
and apply an external perturbation potential which breaks some of original  symmetries explicitly.
Obviously the zero modes associated with the symmetries which the external potential breaks
turn into non-zero ones. The relevant orders are affected substantially in general, and
 in some cases the ordered state may collapse rapidly.
Thus what the zero modes change into when the external perturbation potential is applied is 
a crucial question. We therefore formulate the BdG equations on the first order perturbation 
with respect to the external potential and derive the eigen equations projected out on the zero and
adjoint modes, which determine new eigenvalues of the originally zero and adjoint modes.
For this it is important to notice the singular behaviors of the zero modes
in the naive perturbation expansion. We propose a consistent treatment to avoid the singularity.
This new treatment works well, and it turns out that the eigenvalues 
can become pure imaginary.  The result implies that a new type of
dynamical instability stemming from the zero mode occurs.

Next, as an application of the above general formulation, we consider 
a one-dimensional homogeneous BEC system with a stationary dark soliton
for which we have the $\theta$- and $x$-zero modes and their adjoint ones,
and analyze  changes of the $x$-zero mode and its adjoint one when two types
of external $x$-dependent perturbation potentials are applied.
We show from analytical calculations that they can turn into two pure imaginary modes for some parameter region of the perturbation potentials, 
which means that the system, the soliton in this case, is dynamically unstable.
Then numerical solutions of the TDGP equation show that the soliton moves substantially even for an infinitesimal fluctuation of the center of the soliton.
This gives the dynamical instability stemming from the zero and adjoint
modes which has been unknown, and a new insight into zero modes.

This paper is organized as follows:  The general formulation of the BdG equations 
and their properties are reviewed briefly in Sect.~II.
In Sect.~III, we develop a perturbative expansion for a general BEC system under 
a small external potential, and derive the eigen equations 
closed to the zero and adjoint mode sector. In Sect.~IV, we consider a one-dimensional
homogeneous BEC system with a dark soliton as an example of our formulation since it allows us to
obtain analytical expressions, and show 
that the $x$-zero and its adjoint modes 
turn into two pure imaginary modes under particular perturbation potentials.
Section~V is devoted to summary and concluding remarks.

\section{Bogoliubov-de Gennes equations and zero modes}

For the purpose of our formulations later, a brief review on 
the eigenstates of the BdG equations and the zero modes
is given in this Section.

The dynamical evolution of the condensate is described by the TDGP equation,
\begin{align} \label{TDGP}
i \tpartial \psi(\bx,t) =  \left( -\frac{\nabla^{2}}{2m}  + V(\bx) +g|\psi(\bx,t)|^{2} \right) \psi(\bx,t).
\end{align}
Here $V(\bx)$, $m$ and $g$ are the external potential, the mass of 
a neutral atom and the coupling constant, respectively.
Throughout this paper, we set $\hbar$ to unity.
The spatial integration of the density of the condensate $|\psi(\bx,t)|^2$
is the total number of condensate $N_c$.
The stationary state of the condensate, denoted by $\psi_{\rm st}(\bx,t)=\xi(\bx)e^{-i\mu t}$, obeys the stationary Gross-Pitaevskii (GP) equation,
\begin{align}\label{GP}
 \left( -\frac{\nabla^{2}}{2m} + V(\bx) - \mu +g|\xi(\bx)|^{2} \right) \xi(\bx)=0.
\end{align}
where $\mu$ is the chemical potential.

In order to analyze the stability of the condensate, we consider a small 
fluctuation from the stationary condensate state, $\delta \psi(\bx,t)e^{-i\mu t}$, as
\begin{align}
\psi(\bx,t) &=\psi_{\rm st}(\bx,t)+\delta \psi(\bx,t)e^{-i\mu t} \nonumber \\
&= \Big[ \xi(\bx) + \sum_{\ell}(u_{\ell}(\bx)e^{-i\omega_{\ell}t}
 + v^*_{\ell}(\bx)e^{i\omega_{\ell}^* t}) \Big]e^{-i\mu t}.
\end{align}

Within the  approximation linear with respect to $\delta \psi(\bx,t)$, the amplitudes $u_{\ell}$ and $v_{\ell}$ obey the following equations which are called the BdG equations: 
\begin{align} \label{BdGeq}
		T \by_{\ell}(\bx) = \omega_{\ell} \by_{\ell}(\bx).
\end{align}
Here we have introduced the doublet notation,
\begin{align}
        \by_{\ell}(\bx) = \left(\!\! \begin{array}{c}
				                        u_{\ell}(\bx) \\
				                        v_{\ell}(\bx) \\
			                      \end{array} \!\!\right), \ 
        T = \left(\!\! \begin{array}{cc}
				     \bdgL    &  \bdgM  \\
				     -\bdgM^* &  -\bdgL \\
			       \end{array} \!\!\right) , 
\end{align}
where
\begin{align}
        \bdgL &= -\frac{\nabla^{2}}{2m} + V - \mu +2g|\xi(\bx)|^{2},\\
        \bdgM &= g\xi^{2}(\bx).
\end{align}
Since the matrix $T$ is a non-hermitian operator in general, the eigenvalue $\omega_{\ell}$ can be complex.
If any eigenvalue is complex, we can judge that the stationary condensate is dynamically unstable.

The properties of the matrix $T$ and the BdG eigenfunctions are summarized according to Refs.~\cite{Nakamura, Kobayashi, Mine}:
\begin{align}
 \sigma_{3} T \sigma_{3} &= T^{\dag}, \\
 \sigma_{1} T \sigma_{1} &= -T^{*},
\end{align}
with the Pauli's $i$-th matrix $\sigma_i$.
Because of these symmetric properties, the BdG eigenfunctions have the following features.
(i) The eigenfunctions are orthogonal to each other under the indefinite inner product, 
\begin{align}
(\bs,\bt)=\int d\bx \bs^{\dag}(\bx) \sigma_3 \bt(\bx),
\end{align}
for two doublets, $\bs$ and $\bt$.
(ii) If $\by$ is an eigenfunction belonging to either real or pure imaginary eigenvalue $\omega$ 
(either $\mathrm{Re}[\omega] \not= 0,\ \mathrm{Im}[\omega] = 0$ or 
$\mathrm{Re}[\omega] = 0,\ \mathrm{Im}[\omega] \not= 0$), 
the eigenfunction belonging to $-\omega$, $\sigma_1 \by^*$ for the former or $\by_*$ for the latter, always emerges~\cite{Kobayashi, Mine}.
Thus the eigenfunctions
belonging to real or pure imaginary eigenvalues form pairs.
Otherwise, when $\by$ is an eigenfunction belonging to a complex eigenvalue $\omega$ ($\mathrm{Re}[\omega] \not= 0,\ \mathrm{Im}[\omega]\not= 0$), it and the eigenfunctions $\sigma_1 \by^*,\  \by_*$ and $\sigma_1 \by_*^*$ belonging to $-\omega^* ,\ \omega^*$ and $-\omega$, respectively, make a quartet.
(iii) A squared norm, defined by $\| \bs \|^2 = (\bs,\bs)$,
 is not restricted to be positive, but can also be negative or zero.
When $\by$ is an eigenfunction belonging to a real eigenvalue, 
the squared norm is normalized to $\pm 1$.
Otherwise, when $\by$ is an eigenfunction belonging to a complex eigenvalue
including a pure imaginary one, the squared norm becomes zero.
In this case, $\by$ is ``normalized'' as $(\by,\by_*)=1$.

Finally, we emphasize that the BdG equations have eigenfunctions belonging
 to zero eigenvalues~\cite{Lewenstein,Matsumoto2,Dziarmaga}, $\by_{0,i}$, corresponding to each spontaneously broken symmetry labeled by $i$.
For example, since the BEC system  spontaneously breaks the U(1) global gauge symmetry, the BdG equations always have the following zero eigenfunction~\cite{Lewenstein,Matsumoto2}:
\begin{align}
  \by_{0,\theta}(\bx) \propto  \left(\!\! \begin{array}{c}
				                        \xi(\bx) \\
				                        -\xi^*(\bx) \\
			                      \end{array} \!\!\right).
\end{align}
Because the zero modes are orthogonal to all the eigenfunctions including themselves,
the simple set of the BdG eigenfunctions is not complete.
It is pointed out in Refs.~\cite{Lewenstein,Matsumoto2,Dziarmaga} that
 the adjoint modes, $\by_{-1,i}$, which are defined by 
\begin{align} \label{adjoint eq}
T\by_{-1,i}=I_{i}\by_{0,i}.
\end{align}
and are nonorthogonal to the corresponding zero modes, must be included 
for the completeness of the set of BdG eigenfunctions.
Here $I_{i}$ is the normalization constant to have $(\by_{0,i}, \by_{-1,i})=1$.

\section{Solutions of Gross-Pitaevskii and Bogoliubov-de Gennes equations under external perturbation potential}
We add a potential $\ve \delta V(\bx)$ with a sufficiently small parameter $\ve$ 
to $V({\bm x})$ in Eq.~(\ref{GP}), so that the original translational symmetry is broken explicitly.
Then the order parameter and chemical potential are expanded in the perturbation series, 
\begin{align}
\xi^{\ve}(\bx) &=\xi^{(0)}(\bx)+\ve \xi^{(1)}(\bx)+\cdots, \label{pertxi}\\
\mu^{\ve} &=\mu^{(0)}+\ve \mu^{(1)}+\cdots . \label{pertmu}
\end{align}

Let $\by^{\ve}_{0}(\bx)$ stand for the BdG eigenfunction
which converges to a linear combination of $\by_{0,i}$
 in the unperturbed limit $\ve \rightarrow 0$,
\begin{align} \label{shift BdG}
 T_{\ve} \by^{\ve}_{0}(\bx) = \delta \omega_{0}^{\ve} \by_{0}^{\ve} \, ,
\end{align}
where the eigenvalue $\delta \omega_{0}^{\ve}$ approaches to 0 as $\ve \rightarrow 0$.
In order to solve these equations perturbatively, we develop our calculations as follows:
First we substitute Eqs.~(\ref{pertxi}) and (\ref{pertmu}) into the BdG $T$-matrix,
\begin{align}
T_{\ve}&= T^{(0)} + \ve T^{(1)} + \cdots, \\
& T^{(0)} = \left(\!\! \begin{array}{cc}
				     \bdgL^{(0)}   &  \bdgM^{(0)}  \\
				     -\bdgM^{(0)*} &  -\bdgL^{(0)} \\
			       \end{array} \!\!\right) ,\\ 
&T^{(1)} = \left(\!\! \begin{array}{cc}
				     \bdgL^{(1)}   &  \bdgM^{(1)}  \\
				     -\bdgM^{(1)*} &  -\bdgL^{(1)} \\
			       \end{array} \!\!\right) ,
\end{align}
where
\begin{align}
 \bdgL^{(0)} &=  -\frac{\nabla^{2}}{2m} \!+\! V(\bx) \!-\! 
\mu^{(0)} \!+\! 2g|\xi^{(0)}(\bx)|^2 , \\
 \bdgM^{(0)} &=  g \xi^{(0)2}(\bx) ,\\
 \bdgL^{(1)} &= - \mu^{(1)} +\delta V(\bx) + 2g (\xi^{(0)}(\bx) \xi^{(1)*}(\bx) + \xi^{(0)*}(\bx) \xi^{(1)}(\bx)), \\
 \bdgM^{(1)} &=  2g \xi^{(0)}(\bx) \xi^{(1)}(\bx).
\end{align}
The equation for $\xi^{(1)}(\bx)$ and $\mu^{(1)}$ is explicitly
\begin{align} \label{GP_1}
\bdgL^{(0)} \xi^{(1)}(\bx) + \bdgM^{(0)} \xi^{(1)*}(\bx) = ( \mu^{(1)} - \delta V(\bx) ) \xi^{(0)}(\bx).
\end{align}
For the unperturbed non-zero real eigenvalue,
the perturbed eigenfunction can be expressed in a naive integer power expansion of $\ve$,
which was demonstrated in Ref.~\cite{Nakamura}.
In contrast, we immediately notice that for the unperturbed zero eigenvalue 
the perturbed eigenfunction can not be expanded in the same manner.
That is seen from the fact that the squared norm of $\by_{0}^{\ve}$ belonging to non-zero real 
$\delta \omega_{0}^{\ve}$ is singular with respect to $\ve$.
Actually it is $1$ or $-1$ at $\ve \not= 0$, and $0$ at $\ve = 0$.

The above singular property with respect to $\ve$ demands a new formulation
of the problem. Our new approach to evade the singularity
is that without using a power expansion of $\ve$, we simply
 expand $\by_{0}^{\ve}$ in
 the complete set of the zero-th order BdG eigenfunctions including the adjoint modes as
\begin{align} \label{expansion}
\by_{0}^{\ve}(\bx) 
  =\sum_{\ell=\text{ex.}} \left\{ A^{\ve}_{\ell}\by_{\ell}(\bx) + B^{\ve}_{\ell} 
\sigma_{1} \by^{*}_{\ell}(\bx) \right\}
    + \sum_{i=\text{z.m.}} 
             \left\{ C^{\ve}_{i}\by_{0,i}(\bx) + D^{\ve}_{i}\by_{-1,i}(\bx) \right\},
\end{align} 
where ``ex." and ``z.m." mean the summations over the excitation and zero modes, respectively.
Projecting $\by_{0}^{\ve}$ on $\by_{\ell}$, $\sigma_{1} \by^{*}_{\ell}$, $\by_{-1,i}$ and $\by_{0,i}$, respectively, we 
can confirm 
\begin{align}
A^{\ve}_{\ell} = O\left( \ve \right)\,, \quad 
B^{\ve}_{\ell} = O\left( \ve \right)\,, 
\end{align}
and obtain new eigen equations within the zero mode sector
\begin{align}
\sum_{j=\text{z.m.}} \left\{\ve C^{\ve}_{j} Y_{-1,0}^{(j i)} + D^{\ve}_{j} (I_{j} \delta_{j i} +\ve Y_{-1,-1}^{(j i)} )\right\} 
&= \delta\omega^{\ve}_{0} C^{\ve}_{i} +O(\ve),   \label{secular_1} \\
\sum_{j=\text{z.m.}} \left\{\ve C^{\ve}_{j} Y_{0,0}^{(j i)} +\ve D^{\ve}_{j}Y_{0,-1}^{(j i)} \right\} 
&= \delta\omega^{\ve}_{0} D^{\ve}_{i} +O(\ve) ,                               \label{secular_2} 
\end{align}
where $Y_{n,m}^{(i j)} = (\by_{n,i}, T^{(1)} \by_{m,j})$ for $\ n,\ m=0,-1$.
Since the matrix in Eqs.~(\ref{secular_1}) and (\ref{secular_2})
 are generally non-hermitian, the eigenvalue $\delta\omega^{\ve}_{0}$ can be complex.
It turns out from the symmetric property of the matrix $T^{(1)}$
that $Y_{0,0}^{(i i)}$ and $Y_{-1,-1}^{(i i)} $ are
real and that $Y_{-1,0}^{(i i)} $ is pure imaginary.
Furthermore, it is easy to check that $Y_{0,0}^{(\theta i)} = 0$, and, therefore, 
\begin{align}
	\delta\omega^{\ve}_{0,\theta}=0,
\end{align}
which is natural since the external potential does not break the U(1) gauge symmetry. 

\section{one-dimensional condensate system with dark soliton}

In this section, the formulation in the previous sections is applied to a concrete system. Our
model system is a one-dimensional homogeneous one (thus $V(x)=0$) with a repulsive interaction $g>0$.
We consider a situation with a single dark soliton, and the system acquires the zero mode corresponding 
to a spontaneous symmetry breakdown of the translation other than 
that corresponding to a spontaneous breakdown of the U(1) gauge symmetry. Then we perturb the system 
by $\delta V(x)$.

In order to evaluate the eigenvalues and eigenfunctions originated from zero modes under the perturbation potential, we deal with a single dark soliton.
The dark soliton solution can be obtained by solving the GP equation under the boundary conditions of $\xi(0)=0$ and $\xi(\pm \infty)=\pm \sqrt{n_{c}}$:
\begin{align}
	\xi(x) = \sqrt{n_c} \tanh (\alpha x ),\quad \mu = gn_c\,.
\end{align}
Here $n_c$ is a bulk density of the condensate and 
\begin{align} \label{alpha}
\alpha = \sqrt{m\mu} = \sqrt{mg n_c}\,.
\end{align}
The BdG equations have the following two zero modes
\begin{align} \label{zeromode1}
 \by_{0,\theta}(x) &= \left( \!\!  \begin{array}{c} \tanh(\alpha x) \\ - \! \tanh(\alpha x) \\ \end{array} \!\!  \right), \\
\label{zeromode2} 
 \by_{0,x}(x) &= i\frac{d}{dx} \left( \!  \begin{array}{c} \tanh(\alpha x) \\ \tanh(\alpha x) \\ \end{array} \!  \right).
\end{align}
Their adjoint modes and the constants $I_{i}$ are
\begin{align}
& \by_{-1,\theta}(x) = \frac{1}{2} \left( \!\! \begin{array}{c} 
                                               \tanh (\alpha x) + \alpha x (1-\tanh^2 (\alpha x) ) \\ 
                                               \tanh (\alpha x) + \alpha x (1-\tanh^2 (\alpha x) ) \\ 
                                               \end{array} \!\! \right), \\
& \by_{-1,x}(x) = \frac{i}{4} \left( \!\! \begin{array}{c} 1 \\ -1 \\ \end{array} \!\! \right),\\
&I_{\theta} = \alpha^2, \qquad \ I_{x} = -\frac{gn_c}{4\alpha}.
\end{align}

We apply an external perturbation potential $\ve \delta V(x)$ to the system, breaking
the translational symmetry explicitly but preserving the $U(1)$ gauge one.
In this situation, the matrix eigen equation in Eqs.~(\ref{secular_1}) and (\ref{secular_2}) is reduced to the $2 \times 2$-matrix one involving only the $x$-zero and its adjoint modes,
\begin{align}
\left(\!\! \begin{array}{cc}
	 \ve Y_{-1,0}^{(x x)}       & I_{x} \!+\!  \ve Y_{-1,-1}^{(x x)}\\
	   \ve Y_{0,0}^{(x x)}      &  \ve Y_{0,-1}^{(x x)}  
\end{array} \!\! \right)
\left(\!\! \begin{array}{cc}
       C^{\ve}_{x}\\
       D^{\ve}_{x}
       \end{array} \!\! \right)
\!=\! \delta\omega_{0,x}^{\ve}
\left( \!\! \begin{array}{cc}
       C^{\ve}_{x}\\
       D^{\ve}_{x}
       \end{array} \!\! \right) \!+\! O(\ve). \label{einen3}
\end{align}
Solving this matrix eigen equation, we obtain
\begin{align} \label{eigenvalue of x}
\delta \omega_{0,x}^{\ve} &= \pm\sqrt{ \ve I_{x}Y_{0,0}^{(xx)}} + O(\ve).
\end{align}
Because $I_{x}$ is always negative, the sign of $Y_{0,0}^{(xx)}$ which depends on the perturbation potential determines whether $\delta \omega_{0,x}^{\ve}$ is real or pure imaginary.

When $\delta \omega_{0,x}^{\ve}$ are real, the corresponding eigenfunctions are  
\begin{align} \label{eigenfunction1}
\by^{(\pm),\ve}_{0,x} &\!\!=\!\! \frac{1}{\sqrt{2}} \left\{ \mp \left| \frac{\! \delta \omega^{\ve}_{0,x} }{I_x} \right|^{-\frac{1}{2}} \!\!\!\!\!\! \by_{0,x} + \left| \frac{\! \delta \omega^{\ve}_{0,x} }{I_x} \right|^{\frac{1}{2}} \!\!  \by_{-1,x}   \!\right\} \!+\! O(\ve). 
\end{align}
where the labels $(\pm)$ indicate the signs of $\delta \omega_{0,x}^{\ve}$.
Note that the squared norms of these eigenfunctions should be normalized as either $1$ or $-1$,
and that they are ``anomalous" modes, whose squared norms have opposite signs to their eigenvalues, 
explicitly $\|\by_{0,x}^{(\pm),\ve}\|^2=\mp 1$. The presence of ``anomalous" modes is well known 
in the analysis of solitons with harmonic~\cite{Soliton_Harmonic} 
and double well potentials~\cite{Soliton_DoubleWell1,Soliton_DoubleWell2}, and gives rise to the Landau instability.

When $\delta \omega_{0,x}^{\ve}$ are pure imaginary, the corresponding eigenfunctions are 
\begin{align} 
\by^{(+),\ve}_{0,x} &\!\!=\!\! \frac{1}{\sqrt{2}} \left\{ i \left| \frac{\! \delta \omega^{\ve}_{0,x} }{I_x} \right|^{-\frac{1}{2}} \!\!\!\!\!\! \by_{0,x} + \left| \frac{\! \delta \omega^{\ve}_{0,x} }{I_x} \right|^{\frac{1}{2}} \!\!  \by_{-1,x}   \!\right\} \!+\! O(\ve),\label{eigenfunction2} \\
\by^{(-),\ve}_{0,x} &\!\!=\!\! \frac{1}{\sqrt{2}} \left\{ \left| \frac{\! \delta \omega^{\ve}_{0,x} }{I_x} \right|^{-\frac{1}{2}} \!\!\!\!\!\! \by_{0,x} + i\left| \frac{\! \delta \omega^{\ve}_{0,x} }{I_x} \right|^{\frac{1}{2}} \!\!  \by_{-1,x}   \!\right\} \!+\! O(\ve). \label{eigenfunction3}
\end{align}
Here the orthonormalization conditions are $(\by^{(\pm),\ve}_{0,x},\by^{(\pm),\ve}_{0,x})=0$ and
$(\by^{(+),\ve}_{0,x},\by^{(-),\ve}_{0,x})=1$  according to the general property of 
eigenfunctions belonging to complex eigenvalues~\cite{Mine}. 

Since $\delta \omega^{\ve}_{0,x}$ is in order of $O \left( \ve^{1/2} \right)$, the order of the first terms in Eqs.~(\ref{eigenfunction1}),~(\ref{eigenfunction2}) and~(\ref{eigenfunction3}) is
 $O \left( \ve^{-1/4} \right)$ , and that of the second terms is $O \left( \ve^{1/4} \right)$.
So the profiles of all these eigenfunctions are quite similar to that of 
the $x$-zero mode eigenfunction, and are called the ``ghost" of the zero 
mode in Ref.~\cite{Soliton_Review}.  
The divergences of the first terms in the limit $\ve \rightarrow 0$ reflect the infrared singularity 
of the zero modes.
We note that the coefficients of the first and second terms in
 Eqs.~(\ref{eigenfunction1}), (\ref{eigenfunction2}) and (\ref{eigenfunction3}) 
(which correspond to $C^{\ve}_{x}$ and $D^{\ve}_{x}$ in Eq.~(\ref{expansion})) 
are not of integer power of $\ve$, as it has been pointed out before.

In order to see the implication of the emergence of the pure imaginary eigenvalues
at the classical level,
we suppose the initial condition that the dark soliton is prepared at $x=0$ 
and its center is displaced by an infinitesimal
distance $\delta$ at $t=0$, namely $\psi^\ve(x,0) = \xi(x+\delta)\simeq \xi(x) 
+\delta \frac{d}{dx}\xi(x)$. In the leading order of $\ve$, $\psi^\ve (x,t)$ is written as
 $\xi(x) +\delta \psi(x,t)$, and $\delta \psi(x,t)$ is expanded in the complete set including
$\by^{(+),\ve}_{0,x}$ and $\by^{(-),\ve}_{0,x}$ in Eqs.~(\ref{eigenfunction2}) and (\ref{eigenfunction3}).
After a certain time, the term proportional to $\by^{(+),\ve}_{0,x}$ becomes dominant in the expansion
of $\delta \psi(x,t)$ due to the factor $e^{|\Imag[\delta \omega^{\ve}_{0,x}]| t}$, which gives
\begin{align}
\psi^\ve (x,t)& \simeq \xi(x) +  \delta\, \frac{d \xi(x)}{dx} e^{|\Imag[\delta \omega^{\ve}_{0,x}]| t}
\nonumber \\
& \simeq \xi \left( x + \delta\, e^{|\Imag[\delta \omega^{\ve}_{0,x}]| t}\right) \, .
\end{align}
This consideration implies that when the pure imaginary modes arise from the $x$-zero
and its adjoint modes, the dark soliton starts to move.
This is true however small the symmetry breaking potential $\ve \delta V$ may be, and therefore
one can say that the system is dynamically unstable in the sense that the original stable 
dark soliton can not be sustained although it does not collapse immediately.

In the next two subsections, we concretely consider two types of perturbation potential,
that is, the delta function type and the oscillating one, for both 
the GP equation (\ref{GP_1}) can be solved analytically.

\subsection{Delta function type potential}
Let us consider the delta function type potential at the center of soliton:
 $\delta V(x) = gn_c \delta(\alpha x)$.
Then the GP equation is solved exactly as
\begin{align}
\xi^{\ve}(x) &=\sqrt{n_c} \tanh \left( \alpha x \right),\\
\mu^{\ve} &= \ gn_c.
\end{align}
This means that $\xi^{(1)}(x)=0$ and $\mu^{(1)}=0$, and
\begin{align} \label{dw_DeltaType}
\delta \omega_{0,x}^{\ve} = \pm i \ve^{\frac{1}{2}}\frac{gn_c}{\sqrt{2}}.
\end{align}
Thus the eigenvalues $\delta \omega_{0,x}^{\ve}$ are pure imaginary for any positive 
${gn_c}$. 
As stated above, this result indicates that the system
is dynamically unstable, namely the dark soliton moves away.

\begin{figure}[thbp]
\center
\includegraphics[width=9cm]{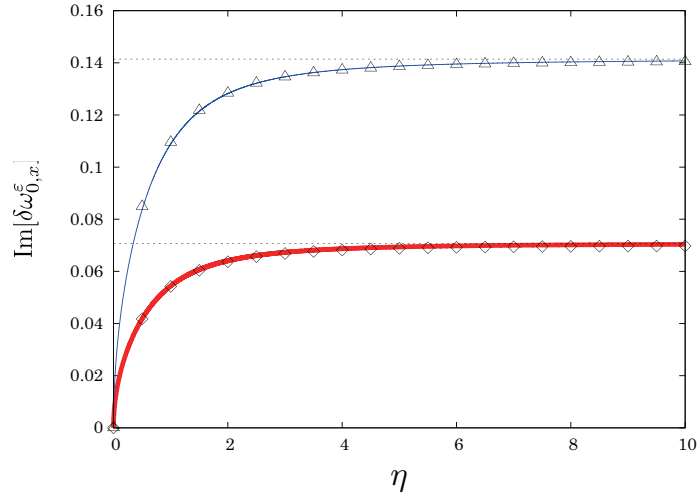}
\caption{\footnotesize{%
The eigenvalue $\delta \omega_{0,x}^{\ve}$ in case of 
the perturbation potential $\delta V_\eta = gn_c 
\eta/\sqrt{\pi}\exp(-\eta^2 (\alpha x)^2)$ with a parameter $\eta$ for $\ve=0.01$. The red (bold) and blue (thin) solid lines indicate the analytic results of perturbation calculation for Eq.~(\ref{eigenvalue of x}) with $gn_c=1$ and $gn_c=2$, respectively, and the open diamonds and triangles indicate the numerical results of 
the BdG equations for $gn_c=1$ and $gn_c=2$.
The two broken lines mean the analytical values in the limit of $\eta\rightarrow\infty$, i.e. Eq.~(\ref{dw_DeltaType}), which are $0.0707\cdots$ for $gn_c=1$ and $0.1414\cdots$ for $gn_c=2$, respectively.
}}
\label{fig:BdG_DeltaType}
\end{figure}

\begin{figure}[thbp]
\center
\includegraphics[width=9cm]{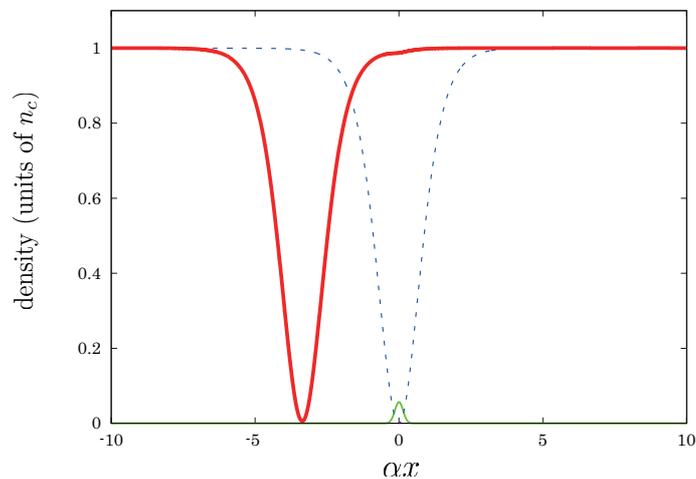}
\caption{\footnotesize{%
The temporal evolution of the soliton obtained by solving the TDGP equation for the perturbation Gaussian potential
$\delta V_\eta$ with the parameters $gn_c=2$, $\eta=5.0$ and $\ve=0.01$.
The blue (broken) line denotes the initial distribution of the soliton whose center is displaced slightly, while the red (solid) one does the distribution after the elapse of a certain time.
The shape of the perturbation potential is depicted in green (thin) line.
}}
\label{fig:TDGP_DeltaType}
\end{figure}

For comparison and in order to avoid the singularity of 
the delta function, we perform numerical calculations for
 the model with a perturbation Gaussian potential
$\delta V_\eta = gn_c \eta/\sqrt{\pi}\exp(-\eta^2  (\alpha x)^2)$ 
with a real parameter $\eta$ which becomes $gn_c \delta(\alpha x)$ as $\eta \rightarrow \infty$.
Figure~\ref{fig:BdG_DeltaType} shows the numerical results
 of the eigenvalues Eq.~(\ref{eigenvalue of x})
for $\delta V_\eta$. 
In Fig.~\ref{fig:TDGP_DeltaType}, we draw a snapshot of the temporal evolution of
 the soliton, obtained by solving the TDGP equation~(\ref{TDGP}) for $\delta V_\eta$. 
It is seen during the time evolution that the condensate is distorted by the potential $\delta V_\eta$ in its vicinity and that accordingly the displacement of the soliton is enhanced.

\subsection{Oscillating type potential}
\begin{figure}[thbp]
\center
\includegraphics[width=7.5cm]{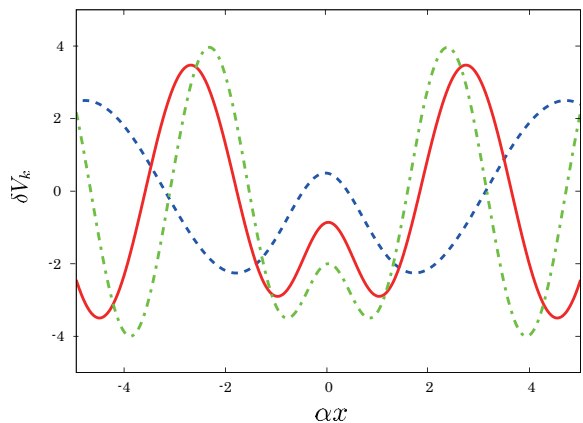}
\caption{\footnotesize{%
The oscillating type potential $\delta V_k(x)$ with the parameter $k=1.0$ (blue broken line), $k=1.73$ (red solid line) and $k=2.0$ (green dashed-dotted line), where $gn_c=1$. 
}}
\label{fig:perturbation potential}
\end{figure}

Suppose the perturbation potential of
\begin{align} \label{OscillatingTypePotential}
\delta V_{k}(x) = gn_{c} \frac{\sin(\alpha k x)}{\tanh(\alpha x)} \left( -3 \tanh^2(\alpha x) +1-\frac{k^2}{2} \right),
\end{align}
with a real parameter $k$.
The shape of this potential is shown in Fig.~\ref{fig:perturbation potential}.
As the parameter $k$ increases, the potential height
 at $x=0$ where the center of the 
stationary soliton is initially placed becomes lower.

\begin{figure}[thb]
\center
\includegraphics[width=7.5cm]{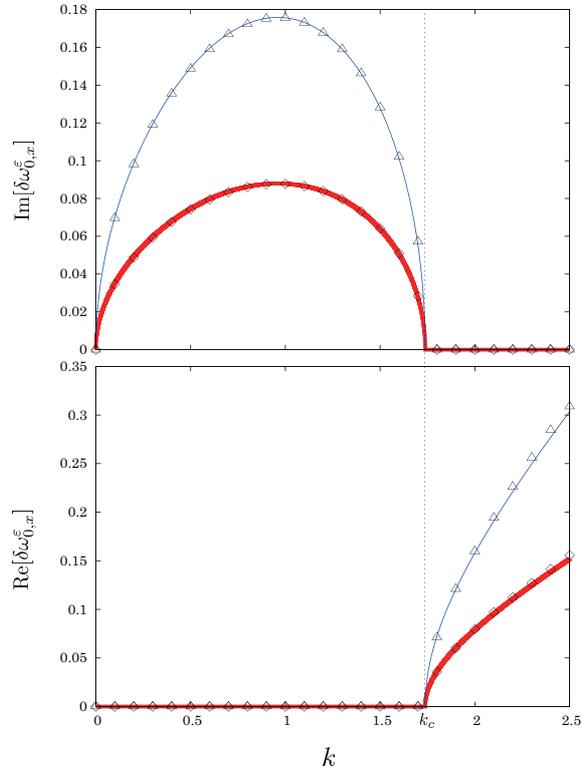}
\caption{\footnotesize{%
The eigenvalue $\delta \omega_{0,x}^{\ve}$ for the perturbation potential
 Eq.~(\ref{OscillatingTypePotential}) with the parameter $k$ for $\ve=0.01$. The red (bold) and blue (thin) solid lines indicate the perturbative results Eq.(\ref{eigenvalue of x}) where $gn_c=1$ and $gn_c=2$, respectively, and the open diamonds and triangles indicate the numerical results of the BdG equations where $gn_c=1$ and $gn_c=2$.
The quantity $k_c=1.73\cdots$ is the critical value of the parameter $k$, 
as is explained in the text.
}}
\label{fig:BdG_OscillatingType}
\end{figure}

The reason why the potential in Eq.~(\ref{OscillatingTypePotential}) is taken is that
one can obtain an analytical solution of the first order GP equation, which is 
\begin{align}
	\xi^{(1)}(x) = \sqrt{n_{c}} \sin(\alpha k x),\quad \mu^{(1)} = 0.
\end{align}
Then $Y^{(xx)}_{0,0}$ is calculated as
\begin{align} \label{Y_xx}
Y^{(xx)}_{0,0}=2 gn_{c} \int ds (1- & \tanh^2(s))^2 \frac{\sin(ks)}{\tanh(s)} \left\{ 3\tanh^2(s) + 1-\frac{k^2}{2} \right\},
\end{align}
where $s=\alpha x$, giving the eigenvalue $\delta \omega_{0,x}^{\ve}$  according to Eq.~(\ref{eigenvalue of x}).
The result is shown in Fig.~\ref{fig:BdG_OscillatingType}. There is a critical value of $k$, denoted by
$k_c=1.73\cdots$, for which $Y^{(xx)}_{0,0}(k_c)=0$.
As $Y^{(xx)}_{0,0}$ is negative for $k>k_c$ and positive for $k<k_c$,
Eq.~(\ref{eigenvalue of x}) yields  real $\delta \omega_{0,x}^{\ve}$ for $k>k_c$
 and pure imaginary one for $k <k_c$, as is
 shown in Fig.~\ref{fig:BdG_OscillatingType}.
Note that the sign of $Y^{(xx)}_{0,0}$
depends on $k$ but not on $\alpha=\sqrt{m g n_c}$, 
so $k_c$ is independent 
of $\alpha$. 
Thus this model exhibits the transition between real and pure imaginary eigenvalues
as the parameter $k$ changes continuously.

Similarly as in the previous subsection, it can be checked from the numerical calculation
of the TDGP equation that in the  region of pure imaginary $\delta \omega_{0,x}^{\ve}$
the dark soliton, displaced very slightly from the center initially, starts to move, and  
oscillates between the two peaks of the potential, that is, the peak at the center and
that next to the center.

\section{Summary}

In this paper, treating the zero modes for the condensate system properly from the viewpoint
of quantum field theory, we have studied possible dynamical instability stemming from the zero modes, and have found eigenfunctions with pure imaginary eigenvalues under a symmetry breaking external perturbation potential.
The pure imaginary modes are non-oscillatory,
and should cause some exponential blowup of small fluctuation or dynamical instability 
in accordance with the general arguments of complex modes. 

As examples, two one-dimensional homogeneous
systems with a dark soliton under perturbation potentials breaking the translational symmetry
have been considered.
Without the perturbation potentials there appears the $x$-zero mode which is the Nambu-Goldstone one
 associated with a spontaneous breakdown of the translational symmetry. 
It is shown that as a result of applying the symmetry breaking potentials it and its adjoint mode can turn into two pure imaginary modes in some cases.  In our case of the soliton,
the relevant transformation is a translation and the dynamical instability manifests itself
such that the center of the soliton moves considerably for its small fluctuation.
We have confirmed this, using two types of the perturbation potentials for
which analytic solutions are available.

The emergence of the pure imaginary modes stemming from the zero modes provides us with a new scenario of that of the complex modes, other than the degeneracy between excitation modes~\cite{Nakamura}.
Although we treat only single-component BECs for simplicity in this paper, the extension to multi-component BEC~\cite{MultiComponent1,MultiComponent2,MultiComponent3} is straightforward.

The arguments about the instability in this paper extend only to the c-number solution of the TDGP equation, but not to quantum fluctuations generated by pure imaginary or complex excitation modes.
It is desirable to take account of the quantum nature of the pure imaginary modes
 properly. To do this, we need to resolve the difficulty that the unperturbed Hamiltonian
for the pure imaginary mode, $H^\ve_{0,{\mathrm{PI}}}$ is not diagonalized and therefore
a unique vacuum is not known \cite{Mine}. Explicitly $H^\ve_{0,{\mathrm{PI}}}$ 
for the one-dimensional soliton model under a symmetry breaking perturbation is given by
\begin{align}
H^\ve_{0,{\mathrm{PI}}}= \delta \omega^\ve_{0,x} b^\dagger a+
\delta \omega^{\ve,\ast}_{0,x} a^\dagger b
\end{align}
with the commutation relations for the operators $a$ and $b$, $[a,b^\dagger]=1$
and $[a,a^\dagger]=[b,b^\dagger]=[a,b]=0$\,.
The diagonalization problem for general complex modes is still open.

In our future work, we will reveal the relation between the dynamical instability in the one-dimensional
solitonic system, presented in this paper, and the snake instability in the two or three-dimensional 
system~\cite{Soliton_Review}.

\section*{Acknowledgements}
This work is partly supported by ``Ambient SoC Global Program of Waseda University" of the Ministry of Education, Culture, 
Sports, Science and Technology, Japan; Grant-in-Aid for Scientific Research (C) (No. 25400410) from the Japan Society for the 
Promotion of Science, Japan;  and Waseda University Grant for Special Research Projects (Project No. 2013B-102). 



\begin{thebibliography}{99}
\bibitem{Cornell}
M.H.~Anderson, J.R.~Ensher, M.R.~Matthews, C.E.~Wieman, and E.A.~Cornell, 
Science 269 (1995) 198.

\bibitem{Ketterle}
K.B.~Davis, M.-O.~Mewes, M.R.~Andrews, N.J.~van Druten, D.S.~Durfee, D.M.~Kurn, and W.~Ketterle, 
Phys.~Rev.~Lett. 75 (1995) 3969.

\bibitem{Bradley}
C.C.~Bradley, C.A.~Sackett, J.J.~Tollett, and R.G.~Hulet, 
Phys.~Rev.~Lett. 75 (1995) 1687.

\bibitem{Shin}
Y.~Shin, M.~Saba, M.~Vengalattore, T.A.~Pasquini, C.~Sanner, A.E.~Leanhardt, M.~Prentiss, D.E.~Pritchard, W.~Ketterle,
Phys.~Rev.~Lett. 93 (2004) 160406.

\bibitem{Fallani}
L.~Fallani, L.~De Sarlo, J.E.~Lye, M.~Modugno, R.~Saers, C.~Fort, and M.~Inguscio
Phys.~Rev.~Lett. 93 (2004) 140406.

\bibitem{Khaykovich}
L.~Khaykovich, F.~Schreck, G.~Ferrari, T.~Bourdel, J.~Cubizolles, L.D.~Carr, Y.~Castin and C.~Salomon,
Science 296 (2002) 1290.

\bibitem{Bogoliubov}
N.N.~Bogoliubov, 
J.~Phys. (Moscow) 11 (1947) 23. 

\bibitem{deGennes}
P.G.~de Gennes, {\it Superconductivity of Metals and Alloys}
(Benjamin, New York, 1966).

\bibitem{Fetter}
A.L.~Fetter, 
Ann.~of Phys. 70 (1972) 67.

\bibitem{Dalfovo}
F.~Dalfovo, S.~Giorgini, L.P.~Pitaevskii and S.~Stringari, 
Rev.~Mod.~Phys. 71 (1999) 463.

\bibitem{Nakamura}
Y.~Nakamura, M.~Mine, M.Okumura, and Y.~Yamanaka, 
Phys.~Rev.~A 77 (2008) 043601.

\bibitem{NGtheorem}
Y.~Nambu and G.~Jona-Lasinio, 
Phys.~Rev. 122 (1961) 345.

\bibitem{Lewenstein}
M.~Lewenstein and L.~You, 
Phys.~Rev.~Lett. 77 (1996) 3489.

\bibitem{Matsumoto2}
H.~Matsumoto and S.~Sakamoto, 
Prog.~Theor.~Phys. 107 (2002) 679.

\bibitem{Dziarmaga}
J.~Dziarmaga, 
Phys.~Rev.~A 70 (2004) 063616.

\bibitem{Kobayashi}
K.~Kobayashi, Y.~Nakamura, M.~Mine, and Y.~Yamanaka,
Ann.~Phys. 324 (2009) 2359.

\bibitem{Mine}
M.~Mine, M.Okumura, T.~Sunaga, and Y.~Yamanaka, 
Ann.~Phys. 322 (2007) 2327.

\bibitem{Soliton_Harmonic}
J.~Dziarmaga and K.~Sacha,
Phys.~Rev.~A 66 (2002) 043620.

\bibitem{Soliton_DoubleWell1}
R.~Ichihara, I.~Danshita and T.~Nikuni,
Phys.~Rev.~A 78 (2008) 063604.

\bibitem{Soliton_DoubleWell2}
S.~Middelkamp, G.~Theocharis, P.G.~Kevrekidis, D.J.~Frantzeskakis, and P.~Schmelcher,
Phys.~Rev.~A 81 (2010) 053618.

\bibitem{Soliton_Review}
D.J.~Frantzeskakis,
J. Phys. A: Math. Theor. 43 (2010) 213001.

\bibitem{MultiComponent1}
S.~Yi, O.~E.~Mustecaplioglu, and L.~You,
Phys.~Rev.~Lett. 90  (2003) 140404.

\bibitem{MultiComponent2}
S.~Yi, O.~E.~Mustecaplioglu, and L.~You,
Phys.~Rev.~A. 68  (2003) 013613.

\bibitem{MultiComponent3}
Y.~Kawaguchi,
Phys.~Rev.~A.  89  (2014) 033627.
\end{thebibliography}
\end{document}